\documentclass[aps,showpacs,twocolumn,superscriptaddress,letterpaper]{revtex4}

\usepackage{graphics}

\begin{document}

\title{A novel FLEX supplemented QMC approach to the Hubbard model}

\author{J.P. Hague} 
\affiliation{Department of Physics, University of Cincinnati}
\affiliation{Computer science and mathematics division, Oak Ridge 
National Laboratory}

\author{Mark Jarrell}
\affiliation{Department of Physics, University of Cincinnati}

\author{T.C.Schulthess}
\affiliation{Computer science and mathematics division, Oak Ridge 
National Laboratory}

\date{\today{}}

\begin{abstract}

This paper introduces a novel ansatz-based technique for solution of
the Hubbard model over two length scales. Short range correlations are
treated exactly using a dynamical cluster approximation QMC
simulation, while longer-length-scale physics requiring larger cluster
sizes is incorporated through the introduction of the fluctuation
exchange (FLEX) approximation. The properties of the resulting hybrid
scheme are examined, and the description of local moment formation is
compared to exact results in 1D. The effects of
electron-electron coupling and electron doping on the shape of the
Fermi-surface are demonstrated in 2D. Causality is examined in both 1D
and 2D. We find that the scheme is successful if QMC clusters of
$N_C\ge 4$ are used (with sufficiently high temperatures in 1D),
however very small QMC clusters of $N_C=1$ lead to acausal results.

\end{abstract}


\maketitle

\section{Introduction}

Despite its transparent meaning and simple form, the Hubbard model is
thought to contain much of the fundamental physics of correlated
electron systems \cite{hubbard1963a}. Since its inception, the Hubbard
model has been successfully applied to the description of the 
Mott--Hubbard
metal-insulator transition, and is thought to contain a minimal
description of certain transition metal oxides, such as the high-$T_C$
cuprates and the low-$T_C$ ruthenates.  While limiting aspects of the
model are easily understood, the exact ground state of the one-band
Hubbard model remains unknown for all cases other than the Hubbard
chain \cite{lieb1968a}.

Examination of the infinite-dimensional limit of the Hubbard model
using dynamical mean-field theory (DMFT) has lead to a greater
understanding of intermediate coupling phenomena such as the
Mott-Hubbard metal-insulator transition \cite{georges1996a}. DMFT maps
the full lattice problem onto a single impurity Anderson model, which
may be solved using various numerically exact techniques such as
quantum Monte-Carlo (QMC) \cite{jarrell1992a} and numerical
renormalization group (NRG) \cite{bulla1999a}. In recent work, the 
dynamical mean-field theory was extended to study correlated electron
systems in 2D, resulting in an approach known as the dynamical cluster
approximation (DCA) \cite{hettler1998a}. DCA is a fully causal method,
which systematically restores non-local corrections to the dynamical
mean-field theory. In previous work, Jarrell \emph{et al.} computed
the phase diagram of the 2D Hubbard model for small clusters of
$N_C=4$ using QMC \cite{jarrell2001a}. Unfortunately, the QMC solution
of large clusters is prohibitively expensive in terms of super-computer 
time,
while alternative approaches, such as the perturbation-theory-based
FLEX approximation can be used to solve large DCA clusters, but only 
in certain limits. FLEX is physically intuitive,
concentrating on scattering from important mechanisms (spin
fluctuation, density fluctuation and particle-particle pairs), and
in the past, FLEX has been shown to reproduce long-length-scale
physics of the Hubbard model such as the Mermin--Wagner theorem
\cite{bickers1989a}.

In this paper we present an ansatz for
combining long length scale information from the FLEX approximation
with the QMC solution of the Hubbard model. The two techniques are
complimentary, since QMC predicts the correct short length scale
physics, while FLEX shows appropriate long length-scale
behavior. In section \ref{sec:formalism}, we introduce the DCA and
the ansatz for the self-energy, and provide technical details of the
modified self-consistent procedure for the incorporation of
two-length-scale physics. We examine issues of causality and 
applicability 
in section \ref{sec:causality}. In section \ref{sec:oned} we examine 
local
moment formation and the momentum dependent occupation
$n_{\mathbf{k}}$. Comparisons are made between the hybrid technique
introduced here, exact results for the 1D model and the more
traditional FLEX technique. We calculate the Fermi surface of the 2D
model in section \ref{sec:twod}. Finally, we discuss the applicability 
and
outlook for the new technique in section \ref{sec:conclusions}

\section{Formalism\label{sec:formalism}}

The DCA extends the DMFT, which assumes that the self-energy is 
constant
across the Brillouin Zone (BZ), by introducing a limited momentum
dependence corresponding to short range spatial fluctuations. This is
achieved by breaking up the Brillouin zone into a series of subzones,
within which it is assumed that the self-energy is constant. This
allows a momentum integrated (or coarse grained) Green function to be
defined in an analogous manner to the definition of DMFT:
\begin{equation}
\bar{G}(i\omega_n,\mathbf{K}_i)=\sum_{k\in
\mathrm{subzone}}\frac{1}{i\omega_n+\mu-\epsilon_k-\Sigma(i\omega_n,\mathbf{K}_i)}
\label{eqn:greensfunction}
\end{equation}
The $\mathbf{K}_i$ are defined as in reference 
\onlinecite{hettler1998a} as 
the mean momentum for a coarse grained cell.  Via an inverse Dyson 
equation, 
this leads to a ``bare'' Green function corresponding to the host of a 
cluster 
impurity problem,
\begin{equation}
\Sigma(\mathbf{K},i\omega_n)={\mathcal{G}}^{-1}_0(\mathbf{K},i\omega_n)-\bar{G}^{-1}(\mathbf{K},i\omega_n)\, 
.
\label{eqn:cegreensfunction}
\end{equation}
which may be solved with a variety of methods.
The approximation has two well defined limiting cases. For cluster 
sizes of 1,
the DCA maps onto the DMFT. For infinite subzones, the formalism is
exact. For a finite number of subzones, results are expected to be
closer to the behavior of the exact infinite lattice than conventional
finite size techniques, since an infinite number of $\mathbf{k}$-states 
is
considered. The coarse graining over $\mathbf{k}$-states maps the
lattice model to be studied (in this case the Hubbard model) onto a
periodic cluster in real space. 

In order to solve the resulting problem, approximations
are employed. Two common cluster solvers are the FLEX approximation
\cite{aryanpor2002a}, and the QMC method of Hirsch and Fye
\cite{jarrell2001b,hirsch1986a}. In principle, the QMC method gives
the exact solution of the cluster, but the numerical cost is high, and
only small clusters can be investigated. Details of the QMC and FLEX 
methods
may be found in references \onlinecite{jarrell1992a} and 
\onlinecite{bickers1989a} respectively.
 
In the hybrid technique proposed here, we define clusters of two sizes, 
one large of size $N_C'$ solved with the FLEX, and one small of size 
$N_C$
solved with QMC.  For the small cluster, coarse grained points are 
denoted 
$\mathbf{K}$, and for the large cluster, they are denoted as 
$\mathbf{K}'$. 
In order to couple the self energies from these methods, we define an 
ansatz
\footnote{The ansatz may be thought of as constructed from a generating 
functional 
$\phi=\phi^{N_C}_{\mathrm{QMC}}-\phi^{N_C}_{\mathrm{FLEX}}+\phi^{N'_C}_{\mathrm{FLEX}}$. From this definition, the self-energy may 
be defined as the first functional derivative, and the irreducible 
vertices as the second functional derivative.}:
\begin{widetext}
\begin{equation} 
\Sigma^{(N_C)}(\mathbf{K},i\omega_n)=\Sigma_{\mathrm{QMC}}^{(N_C)}(\mathbf{K},i\omega_n)-\Sigma_{\mathrm{FLEX}}^{(N_C)}(\mathbf{K},i\omega_n)+\bar{\Sigma}_{\mathrm{FLEX}}^{(N'_C)}(\mathbf{K},i\omega_n)
 \label{eqn:ansatz1}
\end{equation}
\begin{equation}
\Sigma^{(N'_C)}(\mathbf{K'},i\omega_n)=\bar{\Sigma}_{\mathrm{QMC}}^{(N_C)}(\mathbf{K'},i\omega_n)-\bar{\Sigma}_{\mathrm{FLEX}}^{(N_C)}(\mathbf{K'},i\omega_n)+\Sigma_{\mathrm{FLEX}}^{(N'_C)}(\mathbf{K'},i\omega_n)
\label{eqn:ansatz2}
\end{equation}
\end{widetext}
 
The superscripts indicate the size of the cluster. Bars over the
self energy indicate that a secondary coarse graining of the lattice
self energy (linear interpolated self energy) from the other cluster
size has been performed, i.e.
\begin{equation}
\bar{\Sigma}(\mathbf{K'},i\omega_n)=\sum_{k\in K'}
\Sigma(\mathbf{k},i\omega_n)
\end{equation}
where
$\Sigma(\mathbf{k},i\omega_n)$ is the linear interpolation (lattice 
self-energy) of the small cluster (an equivalent expression exists to 
coarse grain from the large to the small cluster).

\begin{figure}
\resizebox{80mm}{!}{\includegraphics{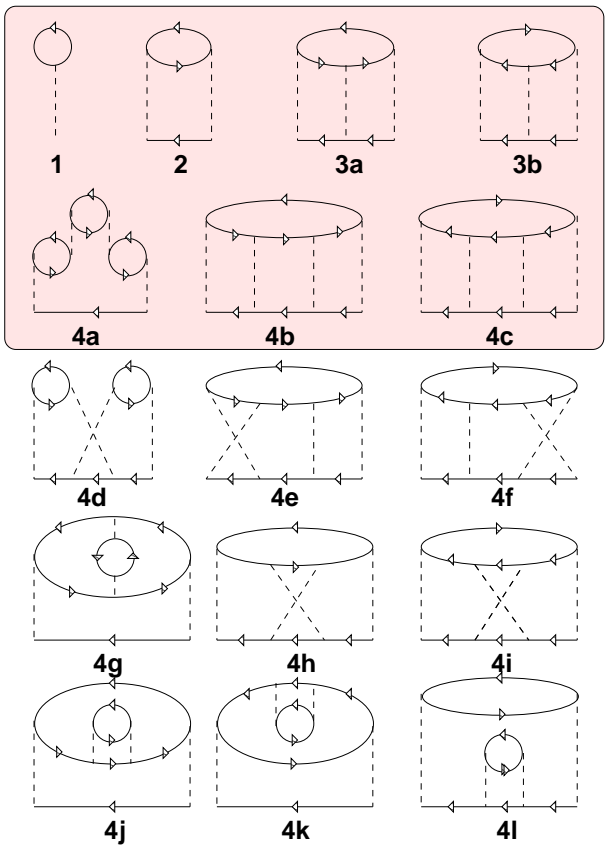}}
\caption{Feynman
diagrams for the Hubbard model to 4th order. Shaded diagrams are
computed for a large cluster, while the remaining diagrams are
computed for the smaller cluster. In our scheme, an infinite number of
FLEX diagrams is considered, and the remaining diagrams are calculated
using QMC, which is a non-perturbative method.}
\label{fig:hub4o}
\end{figure}

The physical content of the
ansatz can be clarified by examining the diagrammatic expansion in
figure \ref{fig:hub4o}. Since the QMC self energy is
non-perturbative, it contains complete contributions from all orders
in $U$ for a cluster of size $N_C$ (1st term). The second term removes
all FLEX contributions for cluster size $N_C$. These are then replaced
with the FLEX contributions for the larger cluster size $N'_C$. This
approach is reasonable if the set of FLEX diagrams have a
stronger momentum dependence than the remaining diagrams. The complete
FLEX with particle-particle scattering obeys the Mermin--Wagner
theorem in 2D, indicating a significant momentum dependence which 
partially 
justifies this assumption.

\begin{figure}
\resizebox{80mm}{!}{\includegraphics{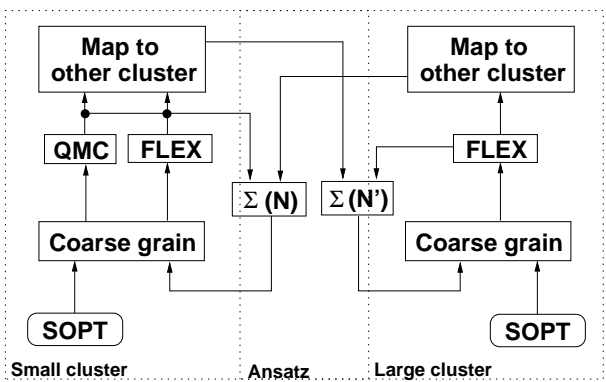}}
\caption{Flow
chart showing the self-consistent procedure for the ansatz. Iteration 
is initialized using second order perturbation theory
(SOPT). The flow then continues with the calculations for both
cluster sizes, which are carried out consecutively. Once convergence is
reached, analysis is carried out on the results from the large
cluster to compute quantities of interest, such as the local moment
and the Fermi-surface.}
\label{fig:flowchart}
\end{figure}

As there are two length scales represented in the
ansatz, two coarse grainings and
cluster exclusions are performed consecutively (one for each length 
scale). The cluster excluded Green
function for the small cluster is then used as input to the QMC, and
the coarse grained Green functions are used as input to the
FLEX. This leads to an iterative procedure which is demonstrated in
figure \ref{fig:flowchart}. There are some additional methods that can 
be employed to aid
convergence. The FLEX self-energy is strongly damped between iterations 
to avoid the FLEX
instability, and many FLEX iterations $\sim 100$ are carried out for
each QMC step (the ansatz is recomputed on each sub-iteration of the 
FLEX). The QMC step is
also damped to a lesser degree.

\section{Range of applicability\label{sec:causality}}

\begin{figure}

\rotatebox{270}{\resizebox{65mm}{85mm}{\includegraphics{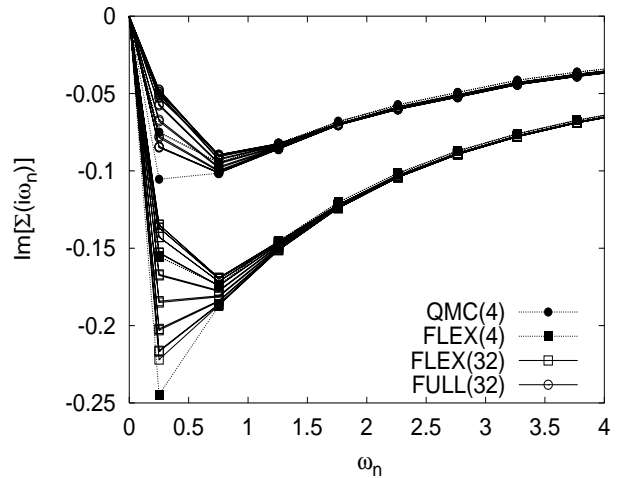}}}

\caption{Self-energy ansatz by contribution, including the QMC part
for the small cluster (filled circles with solid lines), the FLEX
part for the small cluster (filled squares with solid lines), the
FLEX part for the large cluster (open squares with dashed lines),and
the combination of these into the ansatz for the large cluster (open
circles with dashed lines). $U=0.8W$, $T=0.08$, $N_C=4$, $N'_C=32$,
$D=1$, $n=0.5$. For each of the self-energy contributions, there are
a number of curves representing the different contributions at each
momentum point as the Brillouin zone is crossed. Causality is
preserved for this value of $U$. FLEX
overestimates the
self-energy, while QMC has the correct order of
magnitude, but has
insufficient details of the momentum dependence. The ansatz corrects
the momentum
dependence, while keeping the value of the self-energy at the 
appropriate order of
magnitude.}
\label{fig:selfenergyansatz}
\end{figure}

Causality is essential for any predictive theory. Causality violations 
in 
single-particle quantities can lead to negative parts of the 
single-particle 
spectrum and density of states and violations of sum rules. Causality 
violations in two-particle quantities can lead to erroneous predictions 
for 
phase transitions. This section examines how causality problems can 
arise, 
the expected regions of applicability for the new approximation, and 
tactics 
that may be employed to avoid numerical instabilities.
Causality is reflected in the analytic properties of the self energy
and Green functions.  For example, the retarded self energy 
$\Sigma(\mathbf{k},\omega)$ is analytic in the upper half of the 
complex frequency 
plane and $-\frac{1}{\pi} {\rm{Im}}  \Sigma(\mathbf{k},\omega) >0$.  
Since there is a subtraction in the ansatz for the self-energy,
causality problems might be anticipated, since the imaginary part of
the real-frequency self-energy can become positive. 
Here, we search for causality violations in the
Matsubara self energy.
It is related to $\Sigma(\mathbf{k},\omega)$ through
\begin{equation}
\Sigma(\mathbf{k},i\omega_n) = \int d\omega 
\frac
{-\frac{1}{\pi} {\rm{Im}}  \Sigma(\mathbf{k},\omega)
}{i\omega_n - \omega}
\end{equation}
Consequently, $-\omega_n {\rm{Im}}  \Sigma(\mathbf{k},i\omega_n) >0$ as
a consequence of causality.  We employ violations of this inequality as 
a sufficient but not necessary indication of the causality violation of 
our formalism.

In figure \ref{fig:selfenergyansatz}, we examine the self-energy in 1D
for $N_C=4$, $N_C'=32$, $U=0.8W$ and $T=0.08$ at half-filling. All
components of the ansatz are shown: The QMC self-energy for the
smaller cluster (filled circles), the FLEX self-energy for the smaller
cluster (filled squares) and the FLEX self-energy for the larger
cluster (open squares). The result of the ansatz is also shown for the
larger cluster only (open circles). The combination of 1D lattice and
intermediate coupling results in an extreme test case for the ansatz,
and causality can be seen to hold for both large and small clusters.

\begin{figure}
\rotatebox{270}{\resizebox{65mm}{85mm}{\includegraphics{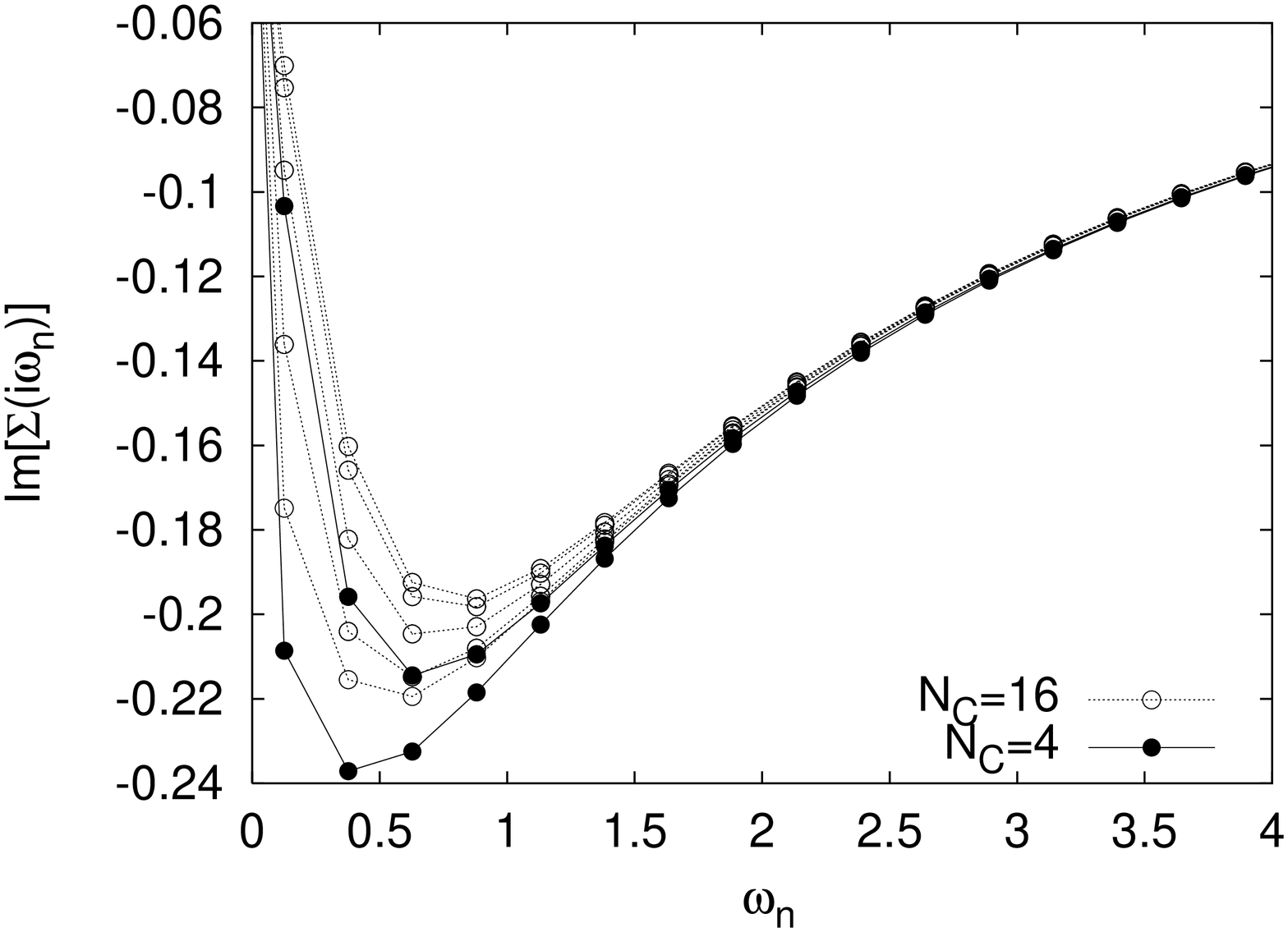}}}
\rotatebox{270}{\resizebox{65mm}{85mm}{\includegraphics{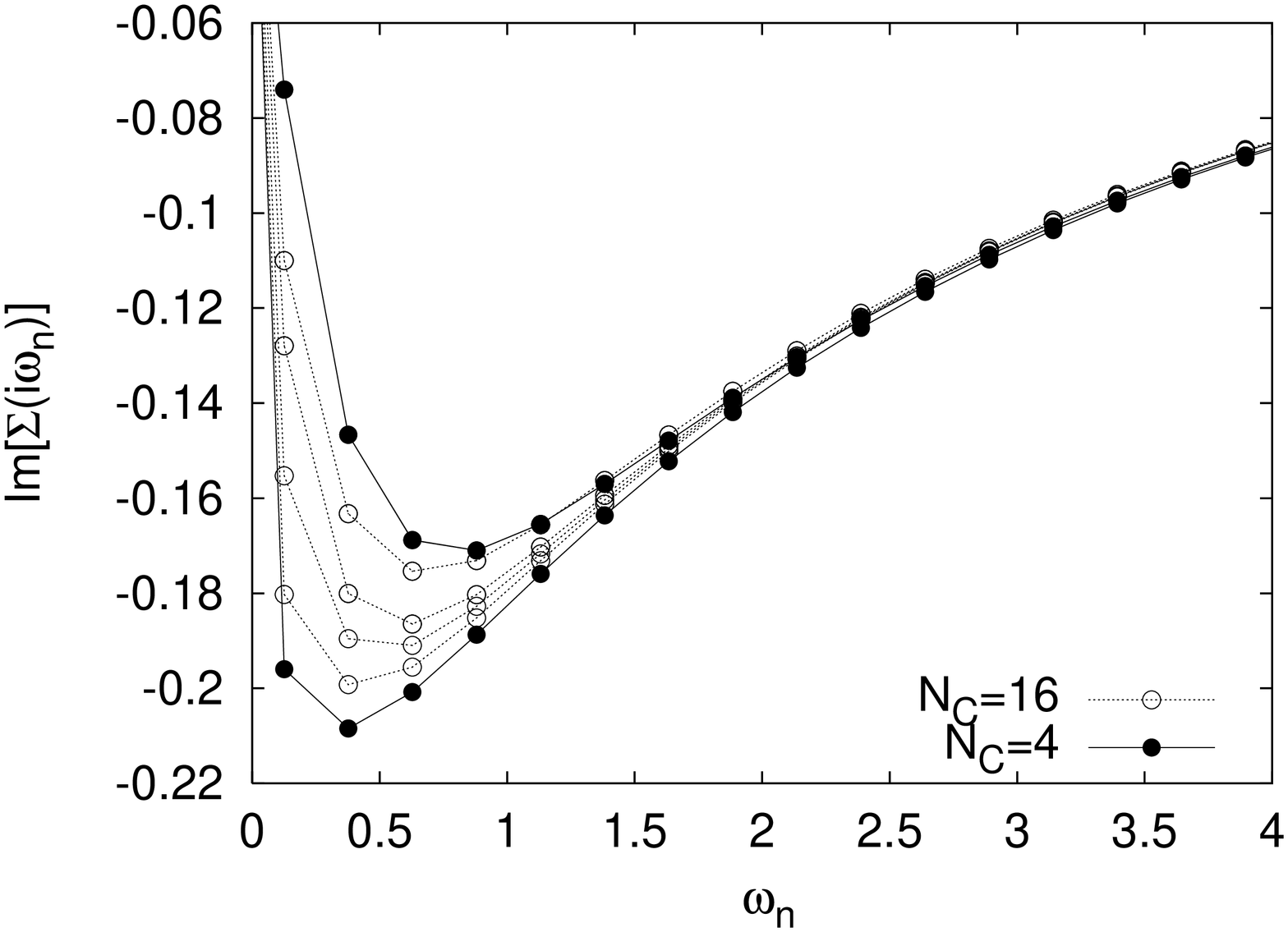}}}
\caption{Matsubara axis self energy calculated using FLEX. The top
 panel shows the self-energy for $U=W$, $D=1$ and the bottom panel for
 $U=0.5W$, $D=2$. In both cases, the result from large and small
 clusters are sufficiently close to one another to avoid causality
 problems in the ansatz.}
\label{fig:selfenergyflex}
\end{figure}

There are two ways in which causality may be violated. The first
relates to the nature of the ansatz. In order to avoid overcounting of
diagrams, FLEX diagrams on the small length scale are subtracted
before new diagrams on a larger length scale are
reinserted. Therefore, even when the FLEX method gives causal results,
the ansatz may return a non-causal self energy. The FLEX parts of the
ansatz are designed to return the correct momentum dependence to the
self energy, and may be seen as a perturbation to the QMC segment of
the self energy,
\begin{equation}
\Sigma(\mathbf{k}) = \Sigma_{\mathrm{QMC}} + \Delta\Sigma(\mathbf{k})
\label{eqn:ansatzrevisited}
\end{equation}

In order to avoid causality problems resulting from the subtraction of
FLEX diagrams, small-length-scale clusters should be as large as
possible, with a minimum requirement that $N_C>1$. The reason for this
is clear for very large QMC clusters, since in the limit that
$N_C\rightarrow N'_C$, both FLEX solvers return the same results, and
$\Delta\Sigma(\mathbf{k})\rightarrow 0$ i.e. causality is
guaranteed. For a good cancellation of the FLEX terms, the small and
large cluster FLEX self-energies should both be on the large N scaling
curve. We demonstrate the scaling behavior of the FLEX approximation
in figure \ref{fig:selfenergyflex} (a pure FLEX calculation is
used). The imaginary part of the self-energy from the FLEX
approximation is shown. Both 1D and 2D calculations are carried out,
and $N_C$ chosen to be representative of cluster sizes used in this
paper. For the 2D case, the $N_C=4$ self-energy doesn't sit directly
on the scaling curve. However, it has the correct form to promote
causality since the extremal behavior is similar to that of the large
cluster. In general, for smaller $U$, less cluster points are needed
to fall on the scaling curve (with a minimum of $N_C=4$ for both 1D
and 2D systems). For larger $U>W$, the minimum $N_C$ required to see
correct scaling behavior in the FLEX approximation is expected to
increase. We suggest that the use of the ansatz should be limited to
couplings that are no bigger than the bandwidth.

If the self-energy contributions from the FLEX approximation lie far
away from the scaling curve, then the magnitude of
$\Delta\Sigma(\mathbf{k})$ can be as large as the QMC part. FLEX
inherently overestimates the magnitude of the self-energy, so errors
are expected to be amplified at strong coupling. For this reason, use
of the DMFT ($N_C=1$) as a small cluster solver is not advised unless 
the temperature is high, or the coupling is weak.

Causality problems may also originate from the FLEX instability. FLEX
is constructed from a geometric series, which imposes a condition on
the susceptibilities:
$\chi_{pp}(\mathbf{0},0),\chi_{ph}(\mathbf{Q},0)<1$. At very high
temperatures, both susceptibilities are small, and the geometric
condition is met. As temperature decreases, the susceptibilities
grow. In 1D and 2D, where the Mermin-Wagner theorem holds, the FLEX
instability should only cause minor issues of numerical instability,
since the susceptibilities are never expected to diverge (N.B. Some
numerical effort is still required to avoid the instability). In 3D,
however, FLEX predicts a phase transition at moderate
temperatures. Below that transition, the unbroken symmetry
approximation is neither causal nor valid. Lower temperatures could
only be accessed by extending the approximation to include the
anomalous Green functions associated with broken symmetry states. It
is unlikely that the Stoner criterion $\chi_{ph}=1$ is meaningful in
the ansatz scheme, so there will be regions of the parameter space
that cannot be reached in 3D. In order to investigate the 3D phase
diagram with this scheme, it is essential that the FLEX instability
occurs at lower temperatures than the true phase
transition. Alternatively a large cluster solver such as
2nd order perturbation theory could to be applied. The phase
transition would be measured through the divergence of the relevant
2-particle susceptibility. To calculate this quantity, it would be
necessary to introduce an additional ansatz for the irreducible vertex
function. This would be quite involved, and is left for a future
publication.

Finally, we note that FLEX (as an extension of the T-matrix
approximation) is exact for dilute systems, and the ansatz is
therefore expected to be exact in the region of low doping. So long as
reasonable temperatures are maintained, results should be applicable
to a wide range of couplings and dopings. For dopings closer to
half-filling, the results of testing show the approximation to be
valid for couplings of up to the band width.

\section{1D Model\label{sec:oned}}

\begin{figure}
\rotatebox{270}{\resizebox{65mm}{85mm}{\includegraphics{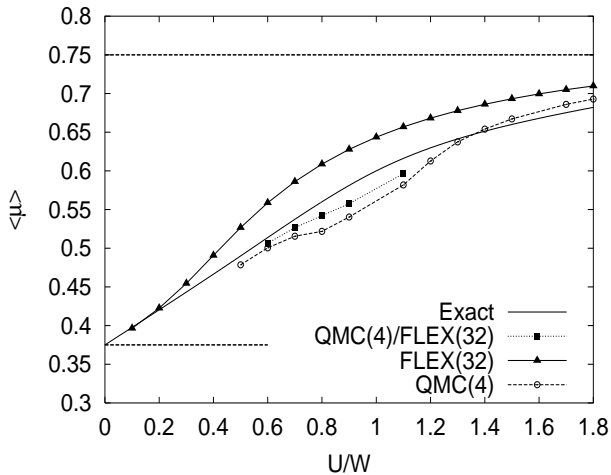}}}
\caption{Comparison of the local moment calculated using FLEX and the
ansatz-based technique presented in this paper against the exact
Bethe-ansatz result. All schemes have the same weak-coupling
behavior. Although FLEX overestimates the moment at strong coupling,
the solution from our ansatz shows promising results. Both the
underestimation of the moment that results from small cluster
simulations, and the overestimation inherent in the FLEX approximation
are corrected.}
\label{fig:localmoment}
\end{figure}

In this section, the applicability of the hybrid FLEX/QMC approach to
the 1D Hubbard model is discussed. Since the exact ground state of the
1D case is known from the Bethe ansatz solution \cite{lieb1968a}, a
quantitative comparison of certain quantities is possible.

While the FLEX approximation predicts the correct AFM transition
temperature (the Mermin--Wagner theorem requires that there is only a
transition at absolute zero), it is well known that FLEX describes
local moment formation incorrectly for couplings of the order of the
bandwidth. The fundamental definition of the local moment is,
\begin{equation}
<\mu>=S(S+1)<(n_{\uparrow}-n_{\downarrow})^2>=\frac{3}{4}(<n>-2<D>)
\end{equation}
where $<D>=<n_{\uparrow}n_{\downarrow}>$ is the expectation value of
the double occupancy. \footnote{Note that in some definitions of the
local moment, the prefactor $S(S+1)$ is omitted.} Since the potential
energy term of the Hubbard model is $U D$, $<D>$ can be extracted
from the expectation value of the potential energy via,
\begin{equation}
<V>=\mathrm{Tr}[\Sigma(\mathbf{K},i\omega_n)G(\mathbf{K},i\omega_n)]=U<D>
\end{equation}

\begin{figure}
\rotatebox{270}{\resizebox{65mm}{85mm}{\includegraphics{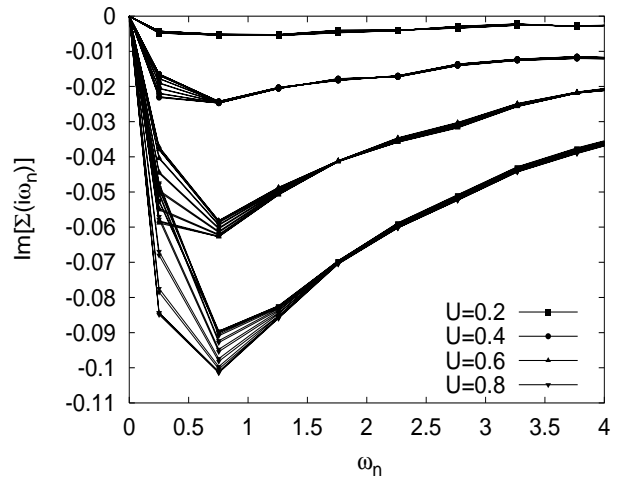}}}
\caption{Self energy in 1D for various couplings. $T=0.08$, $N_C=4$,
$N'_C=32$, $D=1$, $n=0.5$. Each set of curves represents the variation
of the self-energy across the Brillouin zone. Note how the momentum
dependence becomes more pronounced as coupling increases. No causality
problems are found for couplings of similar size to the bandwidth.}
\label{fig:selfenergy1d}
\end{figure}

Figure \ref{fig:localmoment} shows the local moment versus coupling,
calculated using several different techniques (FLEX, the ansatz-based
FLEX/QMC technique presented in this paper, and the exact ground state
solution). There is a difference in temperature between the exact
Bethe-ansatz solution of Lieb and Wu ($T=0$) \cite{lieb1968a} and the
approximate solutions computed here ($T=0.16W$) \footnote{The
relatively high temperature used here avoids instabilities at lower
temperatures}. The temperature is still low enough to see the
non-trivial effects of spatial fluctuations. The hopping, $t=0.25$,
fixes the the bandwidth of the non-interacting problem to unity
(therefore the interaction energy scale for the boundary between weak
and strong coupling regimes is $U=1$). Calculations are carried out
for values of $U$ up to the bandwidth to examine behavior outside the
perturbative regime. The exact result for a 1D problem calculated from
the Bethe ansatz solution is shown, and compared with the 4 site QMC,
the FLEX solution for a cluster size of $N_C=32$, and the hybrid
FLEX/QMC scheme. 

The results in figure \ref{fig:localmoment} shows that all schemes
have the same weak coupling behavior.  The gradient of the low $U$
curve increases with larger cluster size and decreased temperature to
converge on the Bethe-ansatz solution. It can be seen that FLEX
overestimates the moment at strong coupling, i.e. it underestimates
the double occupancy. Alternatively, results from the 4 site QMC
approximation are inclined to underestimate the local moment at
intermediate coupling, probably because the mean-field nature of the
$N_C=4$ DCA used here predicts a metal insulator transition at a
critical coupling $U=U_C$, rather than at $U=0^{+}$ as expected in
1D. The solution from the new ansatz shows promising results. The
moment has the same weak coupling behavior seen in all the presented
approximations. More importantly, the overestimation of the moment
that was predicted by the FLEX approximation has been
corrected. Remarkably, the new hybrid method predicts a moment that
closely follows the exact solution, and the mean-field like behavior
of the DCA is greatly reduced.

The self energy predicted by the hybrid scheme is shown in figure
\ref{fig:selfenergy1d}. Calculations are also carried out for the 1D
system for a series of couplings. In this case, the temperature is
$T=0.08W$. Groups of curves with similar asymptotic behavior belong
to the same coupling, and differ only by their location in the
Brillouin zone. As coupling increases, the self-energy gets larger, as
expected. As compared to a conventional QMC calculation with $N_C=4$,
the self-energy has significantly more detail, representing many more
points in the Brillouin zone. This momentum dependence is most
important for intermediate coupling results.

By examining the $U=0.8W$ result, the advantage of the hybrid method
over the traditional approach can be seen. With a QMC calculation, a
linear interpolation would be used to extract the lattice self
energy. This would mean that all the self-energy curves would be
equidistant. In contract, the lines at the extremes (which correspond
to the center and edge of the Brillouin zone) are clearly much closer
together. In fact, for QMC cluster sizes up to $N_C=6$ in 1D, only
extremal behavior can be predicted, and therefore it is apparent that
the Hybrid scheme generates results which are much closer to the true
lattice self-energy.

\section{2D Model\label{sec:twod}}

\begin{figure}
\resizebox{65mm}{!}{\includegraphics{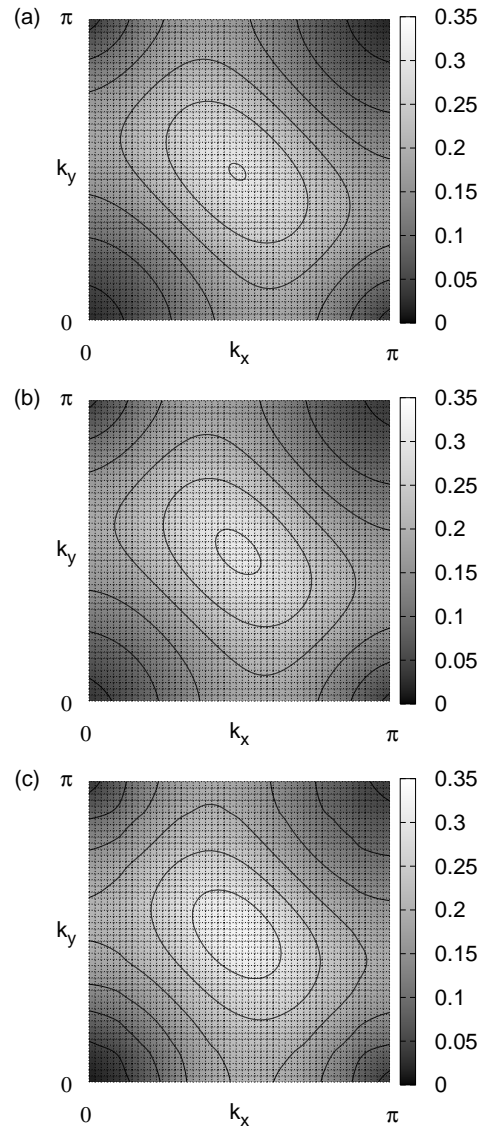}}
\caption{Variation of the Fermi-surface, $|\nabla n(\mathbf{k})|$
with cluster size. Calculations are performed with $U=2$, $T=0.04$,
$\mu=0.075$ and (a) $N'_C=4$, (b) $N'_C=16$ and (c) $N'_C=256$. Larger
cluster sizes lead to a more well defined surface, with a more
electron-like character (see c). Contours are spaced every 
$\Delta|\nabla
n(\mathbf{k})|=0.05$.}
\label{fig:fermisurfacevcluster}
\end{figure} 

\begin{figure}
\resizebox{65mm}{!}{\includegraphics{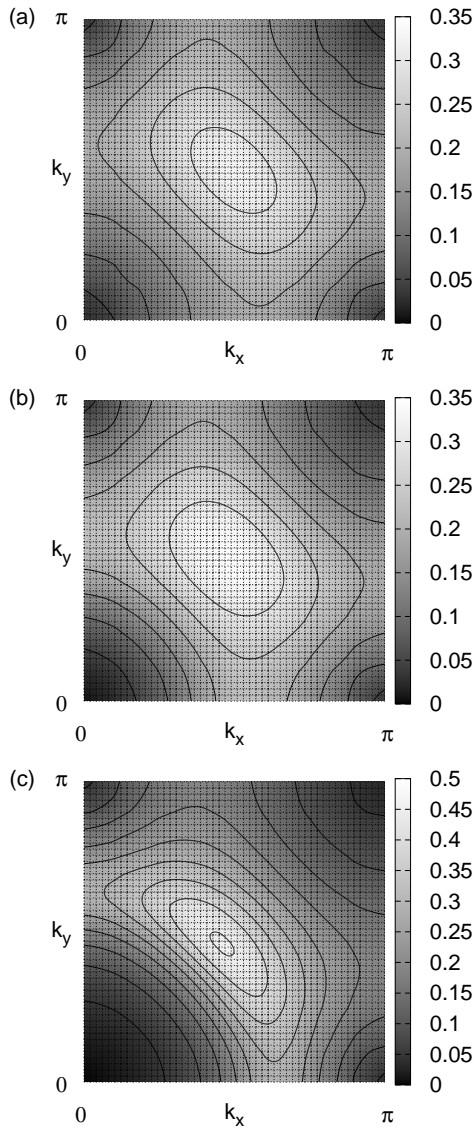}}
\caption{Evolution of the Fermi-surface in 2D, $|\nabla
n(\mathbf{k})|$, for various fillings calculated using the hybrid
FLEX/QMC scheme. All calculations are performed with $U=2=W$,
$T=0.04$, $N_c=4$ and $N'_C=256$. Values of the chemical potential are
(a) $\mu=0.025$, (b) $\mu=0.1$ and (c) $\mu=0.25$. For values of
filling close to half-filling, no Fermi-surface is seen, with a very
diffuse gradient to the momentum dependent electron density. (b) shows
an interim value, where the Fermi-surface begins to form as the edge
of the gap is reached. It can be seen that the central peak develops
greater intensity, and the beginnings of a curved Fermi-surface can be
seen close to the $(\pi/2,\pi/2)$ point. Finally in (c) a clearly
defined Fermi surface is seen, with significant additional intensity. 
Contours are spaced every $\Delta|\nabla
n(\mathbf{k})|=0.05$}
\label{fig:fermisurfacevfilling}
\end{figure}

\begin{figure}
\rotatebox{270}{\resizebox{65mm}{85mm}{\includegraphics{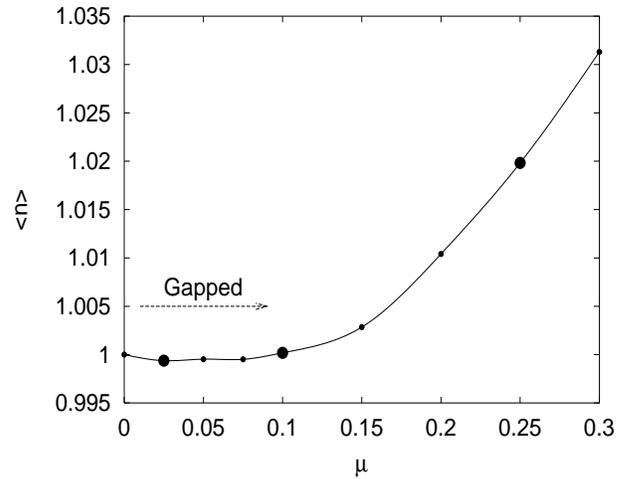}}}
\caption{Evolution of the filling, $<n>$, with chemical potential, 
$\mu$. The signature of a gapped state can be seen at low chemical 
potentials, with no variation of filling as chemical potential is increased. A 
gap of at least $0.05W$ can be seen ($W=2$ in this case). The three 
larger points correspond to the chemical potentials chosen in figure 
\ref{fig:fermisurfacevfilling}. The line is a guide to the eye.}
\label{fig:nvsmu}
\end{figure}

An important quantity that can shed light on the physics of correlated
electron systems is the existence and shape of the Fermi-surface. The
Fermi-surface may be used as input for a variety of other theoretical
techniques, and may be also be compared directly with results from
experimental methods such as angle-resolved photo-emission
spectroscopy (ARPES) and de Haas van Alphen measurements. In this
section, we examine features of the Fermi surface in 2D, by
investigating the momentum-dependent electron density across the
Brillouin zone. The form of this quantity is closely connected to the
shape of the Fermi-surface.

The momentum dependent electron occupation may be calculated using the
following formula,
\begin{equation}
n_{\mathbf{k}}=T\sum_n G(\mathbf{k},i\omega_n) \exp(i\omega_n 0^+)
\label{eqn:gradnk}
\end{equation}
where the lattice Green function has been constructed from the linear
interpolation of the large cluster self-energy. The magnitude of the
gradient of this quantity $|\nabla n_{\mathbf{k}}|$ is related to the
Fermi-surface, since the gradient is largest at the Fermi-surface.

The calculations in this section are performed at $T=0.04$ with
coupling $U=2.0=W$. An $N_C=4$ QMC cluster is used throughout. 48
time slices were used for the Trotter decomposition of the QMC
cluster. The 2D model has a simple cubic tight-binding dispersion with
$t_{x}=t_{y}=0.25$ and a small interplane hopping $t_{z}=0.005$ to
stabilize the solution. The Fermi-surface has been computed using the
hybrid scheme for a series of fillings and large cluster sizes.

One of the underlying aims of the hybrid scheme is to predict features
consistent with very large systems, without the need for expensive QMC
simulations. We first investigate the variation in $|\nabla
n_{\mathbf{k}}|$ as the cluster size is increased from $N'_C=4$ to
$N'_C=256$, when the chemical potential is set to $\mu=0.075$
corresponding to a 2D Hubbard model just away from half-filling. The
results of these calculations are shown in figure
\ref{fig:fermisurfacevcluster}. As cluster size is increased, $|\nabla
n_{\mathbf{k}}|$ becomes sharper and better defined. Also a small
number of finer features emerges. Although $N'_C=64$ is not shown, we
note that the $N'_c=64$ cluster is close to convergence, with only
small differences between $N'_C=64$ and $N'_C=256$. The larger cluster
seems significantly more electron- than hole-like, indicating that the
band-gap inherent at half filling has become slightly narrower,
although the diffuse nature of the surface indicates that the system
is not in a Fermi-liquid state (this is to be expected close to
half-filling, since there are significant spin-fluctuations).

An aspect of the 2D Hubbard model that is of general interest is the
evolution of the Fermi-surface as filling is changed. In a dilute
system, the Hubbard model is well described by the T-matrix
approximation, which predicts Fermi-liquid behavior. At the other
extreme, the half-filled system has a metal-insulator transition, with
strong spin-fluctuations due to the proximity of a phase transition to
the anti-ferromagnetic state at absolute zero. Also, non-Fermi-liquid
pseudogapped behavior has been reported for the 2D Hubbard model a
little off half filling.

In figure \ref{fig:fermisurfacevfilling}, $|\nabla n_{\mathbf{k}}|$ is
shown for the $N_C=4$, $N'_C=256$ scheme. The results are computed
for a fixed $U$ and temperature value, and only the filling is
varied. Representative graphs are shown to demonstrate the gapped
state, the movement between gapped and metallic states, and the
beginnings of the Fermi-surface.

When close to half of the electronic states are occupied $<n>=1.0$
(see the result for $\mu=0.025$), $|\nabla n_{\mathbf{k}}|$ is quite
disperse, with available electronic states across the Brillouin zone,
and no clearly defined maximum associated with a Fermi-surface. As
filling is increased to $<n>=1.002$ ($\mu=0.1$), the Fermi-surface
becomes better defined, and some curvature due to free electrons is
evident. For larger fillings, $<n>=1.02$ ($\mu=0.25$), the
beginnings of an electronic Fermi-surface are evident. In addition to
the electron-like states, an unusual hole-like surface can be
seen. This is smaller in magnitude than the maximum corresponding to
the electronic states, but it persists into the metallic state.

In order to quantify the transition between gapped and metallic
states, the expectation value of the occupation, $<n>$, has been
calculated as a function of the chemical potential, $\mu$. The results
are shown in figure \ref{fig:nvsmu}. For chemical potentials of up to
the order of $\mu=0.1$, the filling does not vary, sitting at
$<n>=1.0$ or half-filling (there is a slight error due to the QMC
algorithm). This indicates the presence of a gap which is approximately
$5\%$ of the band width. Once the metallic regime is reached, the
filling doesn't vary very quickly, which suggests that there are very
few states close to the Fermi-energy. This may be an indication of
pseudogapped behavior.

\section{Conclusions\label{sec:conclusions}}

We have introduced a new technique for the simulation of the Hubbard
model. In this technique, short length-scale fluctuations are treated
using QMC techniques, and the solution is supplemented with long
length scale physics from the FLEX approximation.

We have demonstrated that the new technique is successful at
simulating the Hubbard model in 1D and 2D, provided that the QMC
cluster is sufficiently large. In particular, we find that results for
the 1D Hubbard model at low temperatures are very close to the exact
Bethe ansatz solution. The new hybrid method works well for couplings
up to the order of the band-width. This is important, since spatial
fluctuations on all length scales are expected to make a major
contribution to the physics of the 1D model. It is therefore expected
that the hybrid QMC/FLEX calculations will give important insight into
the 1D problem, while remaining numerically cheap. In 2D, the ansatz
has been applied to the calculation of the Fermi-surface. We
demonstrate that very large clusters of $N'_C=256$ can be
simulated. For coupling of the order of the band-width, we demonstrate
the evolution from insulating to metallic behavior.

The new technique is important from two view points. First, QMC
simulation of large clusters is very expensive, with computing time
growing quickly with cluster size. Therefore, any improvement on the
convergence properties of the DCA method will aid in the accurate
simulation of lattice models. Second, it has been proposed that a
similar technique be used to combine DMFT with the \emph{ab-initio} GW
approximation \cite{sun2002a}. In fact the two cluster solver approach
is quite general, and it is expected that other models could be
accurately treated using similar techniques.

\section{Acknowledgements}
We acknowledge useful conversations with K.\ Aryanpour, D. Hess, 
H.R.\ Krishnamurthy, and Th.\ Maier. This research was supported 
by the NSF grant DMR-0312680, and the Division of Materials Science and 
Engineering, U. S. Department of Energy under Contract 
DE-AC05-00OR22725 
with UT-Battelle, LLC.

\bibliography{dcaqmcflex}

\end{document}